\documentclass[aps,prl,reprint]{revtex4-1}

\usepackage{amsmath}    
\usepackage{graphicx}   
\usepackage{color}      
\usepackage{subfigure}  
\usepackage{hyperref}   

\usepackage{subfigure}
\usepackage{graphicx}
\usepackage{color}
\usepackage[usenames,dvipsnames,svgnames,table]{xcolor}
\usepackage{scalefnt}
\usepackage{setspace}
\usepackage{braket}
\usepackage{amssymb}
\usepackage{siunitx}

\newcommand{\be}{\begin{equation}}
\newcommand{\ee}{\end{equation}}

\begin{document}

\title{Nematic correlation length in iron-based superconductors probed by inelastic x-ray scattering}

\author{A. M. Merritt$^1$, F. Weber$^{2,3}$, J.-P. Castellan$^{2,4}$, Th. Wolf$^2$, D. Ishikawa$^5$, A. H. Said$^6$, A. Alatas$^6$, R. M. Fernandes$^7$, A. Q. R. Baron$^5$, D. Reznik$^{1,8*}$}

\affiliation{1. Department of Physics, University of Colorado at Boulder, Boulder, Colorado 80309, USA\\2. Institute for Solid State Physics, Karlsruhe Institute of Technology, 76021 Karlsruhe, Germany\\3.Institute for Quantum Materials and Technologies, Karlsruhe Institute of Technology, 76021 Karlsruhe, Germany\\4.CEA Saclay, Laboratoire L\'eon Brillouin, F-91191 Gif sur Yvette, France\\5. Materials Dynamics Laboratory, RIKEN SPring-8 Center, RIKEN, 1-1-1 Kouto, Sayo, Hyogo 679-5148 Japan\\6. Advanced Photon Source, Argonne National Laboratory, Argonne, Illinois 60439, USA\\7. School of Physics and Astronomy, University of Minnesota, Minneapolis, Minnesota 55455, USA\\8. Center for Experiments on Quantum Materials, University of Colorado at Boulder, Boulder, Colorado 80309, USA\\* Corresponding author: Dmitry.Reznik@colorado.edu}

\begin{abstract}
Nematicity is ubiquitous in electronic phases of high-$T_c$ superconductors, particularly in the Fe-based systems. We used inelastic x-ray scattering to extract the temperature-dependent nematic correlation length $\xi$ from the anomalous softening of acoustic phonon modes in FeSe, underdoped Ba(Fe$_{0.97}$Co$_{0.03}$)$_2$As$_2$ and optimally doped Ba(Fe$_{0.94}$Co$_{0.06}$)$_2$As$_2$. In all cases, we find that $\xi$ is well described by a power law $(T-T_0)^{-1/2}$ extending over a wide temperature range. Combined with the previously reported Curie-Weiss behavior of the nematic susceptibility, these results point to the mean-field character of the nematic transition, which we attribute to a sizable nemato-elastic coupling that is likely detrimental to superconductivity.
\end{abstract}

\maketitle

The lowering of a high-temperature crystal structure symmetry from tetragonal (fourfold) to orthorhombic (twofold) can be driven by a lattice instability, by a density-wave, or by electronic correlations. In the latter case, since translational symmetry is preserved, the orthorhombic phase is called nematic, in analogy with liquid crystals \cite{Fradkin_NematicFermiFluids-2010}. Even though in this case the lattice is not the driving force behind nematicity, it responds to nematic order and nematic fluctuations due to the coupling to the electronic degrees of freedom \cite{Bohmer_Electronicnematicsusceptibility-2016, fernandes_what_2014}. Indeed, the lowering of the symmetry of the electronic state from fourfold to twofold leads to an orthorhombic atomic lattice distortion, while nematic fluctuations soften the relevant elastic constants \cite{Fernandes10}. In many Fe-based superconductors, such as doped BaFe$_2$As$_2$, nematicity is believed to arise as a vestigial order of the stripe spin-density wave state that sets in at a lower temperature and selects one of two orthogonal wave-vectors related by a $90^{\circ}$ rotation \cite{Fang_Theoryelectronnematic-2008, Xu_Isingspinorders-2008, Paglione_Hightemperaturesuperconductivityironbased-2010, Fernandes_Preemptivenematicorder-2012}. An exception may be FeSe, where nematic order sets in at 90K, but magnetic order does not form at any temperature at ambient pressure \cite{mcqueen_tetragonal--orthorhombic_2009, Mizuguchi_ReviewFeChalcogenides-2010, doi:10.1080/00018732.2010.513480}, although antiferromagnetic (AFM) order appears under pressure \cite{Bohmer_Distinctpressureevolution-2019a}. The origin of nematic order in FeSe remains a topic of intense debate \cite{Baek2015,Bohmer_OriginTetragonaltoOrthorhombicPhase-2015,Wang2016,Fanfarillo2016,chen_anisotropic_2019}.


The impact of the electron-phonon coupling on the electronic orders of Fe-based superconductors has been investigated in different contexts. Density functional theory predicts weak coupling of phonons to electronic charge fluctuations, but significant magnetoelastic coupling of some optic phonons \cite{Boeri_LaFeAsOF-2008, Yin_ElectronHoleSymmetryMagnetic-2008}. Experiments showed weaker effects but agreed qualitatively with these predictions \cite{reznik_phonons_2009,Murai_Effectmagnetismlattice-2016}. Transverse acoustic (TA) phonons dispersing in the [100] direction exhibit the strongest experimentally observed electron-phonon coupling. They soften with temperature ($T$) on approach to the orthorhombic distortion of the atomic lattice in the nematic phase \cite{niedziela_phonon_2011}. Quantitative analysis of this softening allows extracting the nematic correlation length $\xi$ \cite{Weber_Softphononsreveal-2018}. In optimally-doped Ba(Fe$_{0.94}$Co$_{0.06}$)$_2$As$_2$ $\xi$ increases upon cooling in the tetragonal phase but is suppressed inside the superconducting phase of the optimally doped compound \cite{Weber_Softphononsreveal-2018}. However, in previous work only small reduced wavevector ($q$) phonons were considered and due to the tilted and broad resolution ellipsoid of the neutron scattering experiments, a more quantitative analysis of the $T$ dependence of $\xi$ was not possible. Furthermore, that study focused only on one compound, not addressing universality of the observed behavior.

Here we compare the T-dependence of $\xi$ in FeSe and underdoped Ba(Fe$_{1-x}$Co$_{x}$)$_2$As$_2$ (UD Ba-122), whose doping level ($x=0.03$) was chosen such that its structural transition temperature $T_S$ was close to that of FeSe. In addition, we performed detailed measurements of an optimally-doped Ba(Fe$_{0.94}$Co$_{0.06}$)$_2$As$_2$ (OP Ba-122) sample, reaching larger wave-vectors than in the previous study. To achieve better wave-vector resolution with larger scattering intensity, we used inelastic x-ray scattering instead of neutron scattering. We find a striking similarity between all three compounds, despite their rather different ground states. Most importantly, we find that the T-dependence of $\xi$ in FeSe and underdoped and optimally doped Ba(Fe$_{1-x}$Co$_{x}$)$_2$As$_2$ is very well described by $(T-T_0)^{-1/2}$. Combined with the Curie-Weiss behavior observed in $\chi_{\mathrm{nem}}$, our results point to a mean-field behavior with fluctuations extending to rather high temperatures above the structural transition temperature, $T_S$. We attribute this mean-field behavior to the coupling to the lattice, which is known theoretically to change the universality class of the nematic transition from Ising-like to mean-field due to the long-range nematic interactions mediated by strain fluctuations. The implications of our results for the emergence of superconductivity are discussed.

Measurements were performed on the RIKEN BL43LXU beamline at SPring-8, Japan \cite{baron2010riken} and on the 30-ID HERIX beamline at the Advanced Photon Source (APS), Argonne National Laboratory, USA \cite{Toellner:mo5010, Said:co5010,0953-8984-13-34-305}. At SPring-8 the photon energy used was 21.747 keV, while at APS the photon energy was 23.724 keV. A 2-Dimensional analyzer array at BL43LXU allowed parallel measurement of multiple transverse momentum transfers (see discussions in Refs. \cite{Baron_HighResolutionInelasticXRay-2019, Baron_2010},  see also Ref. \cite{baron_introduction_2015}). See supplementary material for further experimental details.

\begin{figure}[htb!]
\includegraphics[width=\linewidth]{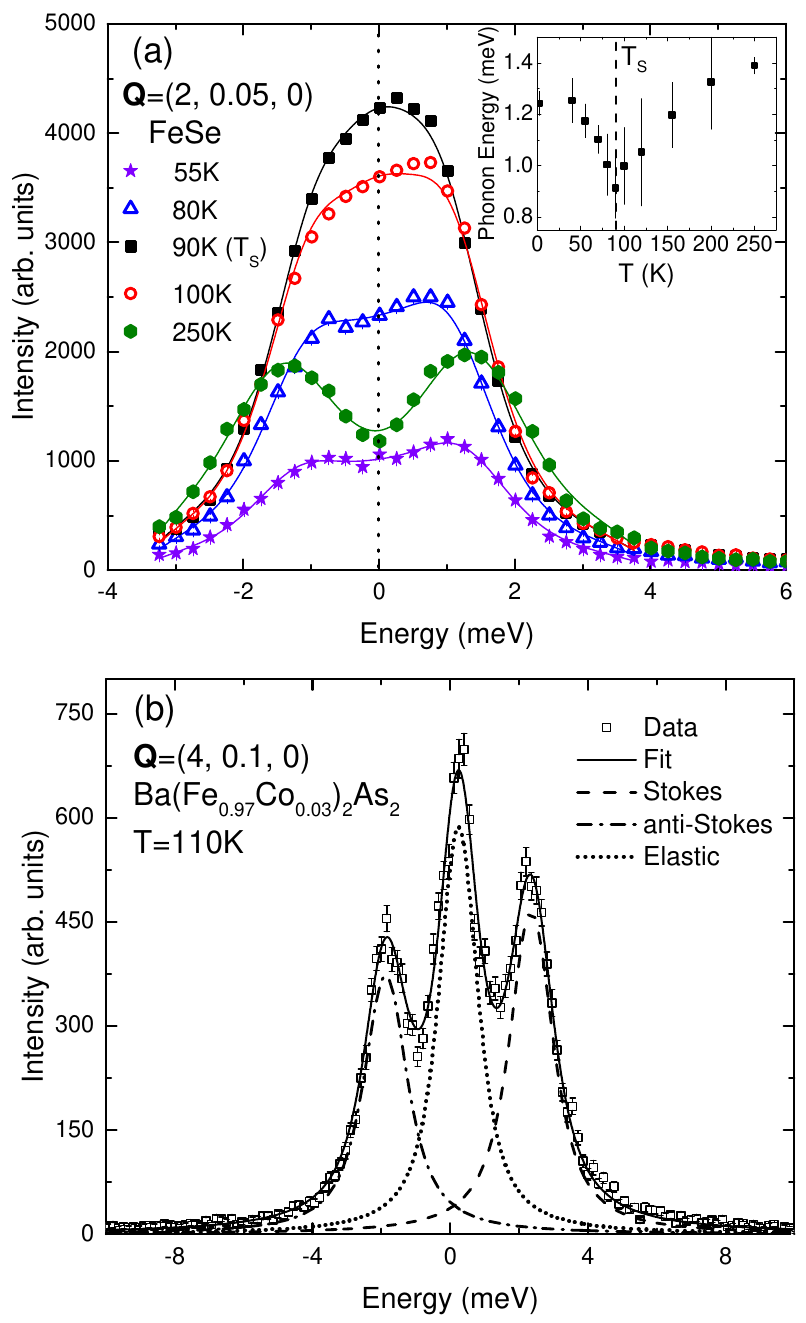}
\caption{Raw data with fits. (a) Energy scans on FeSe at $\textbf{Q}=(2, 0.05, 0)$. Data taken at $T_S=90K$ are represented by the black squares. Error bars are similar in size to the symbols. Inset: phonon energy at $\textbf{Q}=(2, 0.05, 0)$, with $T_S$ marked by the  dashed line. (b) An example fit for data on Ba(Fe$_{0.97}$Co$_{0.03}$)$_2$As$_2$ at $\textbf{Q}=(4, 0.1, 0)$, $T=110K$. The raw data is represented by the empty symbols, the total fit by the solid black line, and then the elastic, Stokes and anti-Stokes peaks by the dotted, dashed and dash-dotted lines, respectively.}
\label{fig:Raw_and_Fit_Data}
\end{figure}

The phonon softening is clearly seen in figure 1a as the separation between energy loss and energy gain peaks decreases and the intensity increases upon cooling towards $T_S$. The trend reverses upon further cooling. For quantitative analysis the three free fit parameters were: The phonon peak intensity, the elastic intensity (not shown and not used in our analysis), and the phonon energy. At $k<0.05$ the phonon intensity was fixed by taking the intensity at the same temperature at $k=0.1$ and using the relationship that the Bose factor corrected intensity of small $q$ acoustic phonons is inversely proportional to the phonon energy \cite{shirane_neutron_2002}. This left only two fitting parameters, which allowed us to fit low $q$ data where the peaks are not visibly separated. Fig. \ref{fig:Raw_and_Fit_Data} shows examples of overall fits at each temperature. Figure \ref{fig:Raw_and_Fit_Data}b, where the peaks are well-separated, shows the individual contributions of the elastic peak plus the Stokes and anti-Stokes phonon peaks. The phonon energy at $\mathbf{Q}=(2, 0.05, 0)$ in FeSe as a function of temperature is similar to the expected behavior of the shear modulus $C_{66}$ from mean field theory \cite{Bohmer_Electronicnematicsusceptibility-2016} (inset of Fig. \ref{fig:Raw_and_Fit_Data}a), which cannot otherwise be observed below $T_S$ by 3-point bending or resonant ultrasound experiments due to twinning in the sample.

As shown in Ref. \cite{Weber_Softphononsreveal-2018}, the phonon energy as a function of momentum transfer, $E(q)$, is related to the nematic correlation length $\xi$ according to:

\be
E(q) = f(q) \sqrt{\frac{C^0_{66}\left(1+\xi^2 q^2 \right)}{\rho\left(\frac{C^0_{66}}{C_{66}}+\xi^2 q^2\right)}}.
\label{fit_equation}
\ee

Here, $\rho$ is the density of the material, the bare shear modulus is given by $C^0_{66}$ and the renormalized shear modulus by $C_{66}$. The latter is related to the former according to $C^0_{66}/C_{66} = 1 + \lambda^2  \chi_{\mathrm{nem}}/C^0_{66}$,
where $\lambda$ is the nemato-elastic coupling constant and $\chi_{\mathrm{nem}}$ is the uniform (i.e. $q=0$) nematic susceptibility \cite{Fernandes10}. The function $f(q)$ is the unrenormalized dispersion, which must vanish linearly with $q$ as $q\rightarrow0$. To fit the data over a wider region of the Brillouin zone, we here use the phenomenological form $f(q) = \lvert \frac{\text{sin}(D q \pi)}{D\pi} \rvert$. The fitting parameter $D$ controls the periodicity of the sine function used for a generic acoustic phonon dispersion. It is fixed by fitting the dispersion at high temperature, where there is little $q$-dependent phonon softening. In the long wavelength limit, $f(q \rightarrow 0) = |q|$, as used in Ref. \cite{Weber_Softphononsreveal-2018}.
 
For our fitting procedure, the renormalized shear modulus, $C_{66}$ is taken from previously reported Young's modulus $Y_{[110]}$ normalized to its high temperature value at 250 K (FeSe) and 293 K (UD Ba-122) \cite{Bohmer_OriginTetragonaltoOrthorhombicPhase-2015}. The temperature dependence of $Y_{[110]}$ is dominated by that of $C_{66}$ if the latter is small, which is the case near the nematic-structural phase transition \cite{bohmer_nematic_2014}. 

\begin{figure}
\includegraphics[width=\linewidth]{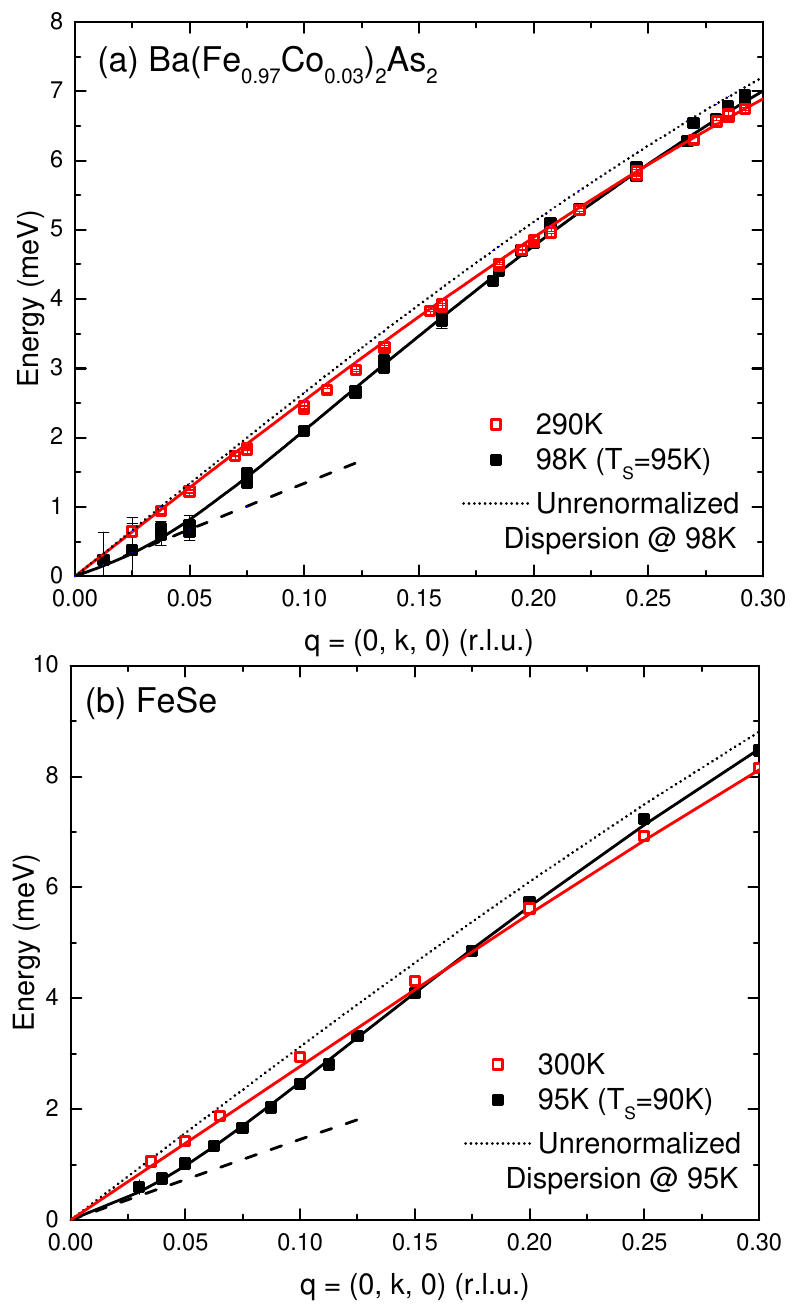}
\caption{Phonon dispersion fits for Ba(Fe$_{0.97}$Co$_{0.03}$)$_2$As$_2$ (a) and FeSe (b). The dotted black line is the expected dispersion in the absence of nematic fluctuations. The data (solid black squares) and fit (solid black curve) show clearly visible softening that increases at low $q$. The dashed line shows the expected low-$q$ slope if the nematic correlation length was very small; it matches the phonon energies only at very low $q$. Hollow red squares/solid red line show data/fit at high temperature respectively.}
\label{fig:Dispersion_Fit}
\end{figure}

Extracting the bare shear modulus $C^0_{66}$ is more complicated, because it requires a complete absence of nematic fluctuations, which is rarely the case in samples displaying a structural transition. Indeed, in SrFe$_2$As$_2$, lattice softening closely match magnetic fluctuations, which persist well above $T_S$ \cite{parshall_close_2015}. Similarly, the measured shear modulus in Ba(Fe$_{1-x}$Co$_{x}$)$_2$As$_2$ varies significantly with doping \cite{Yoshizawa_StructuralQuantumCriticality-2012}. From these observations we conclude that nematic fluctuations in both materials may significantly affect the phonon energy even at room temperature.

\begin{figure}
\includegraphics[width=\linewidth]{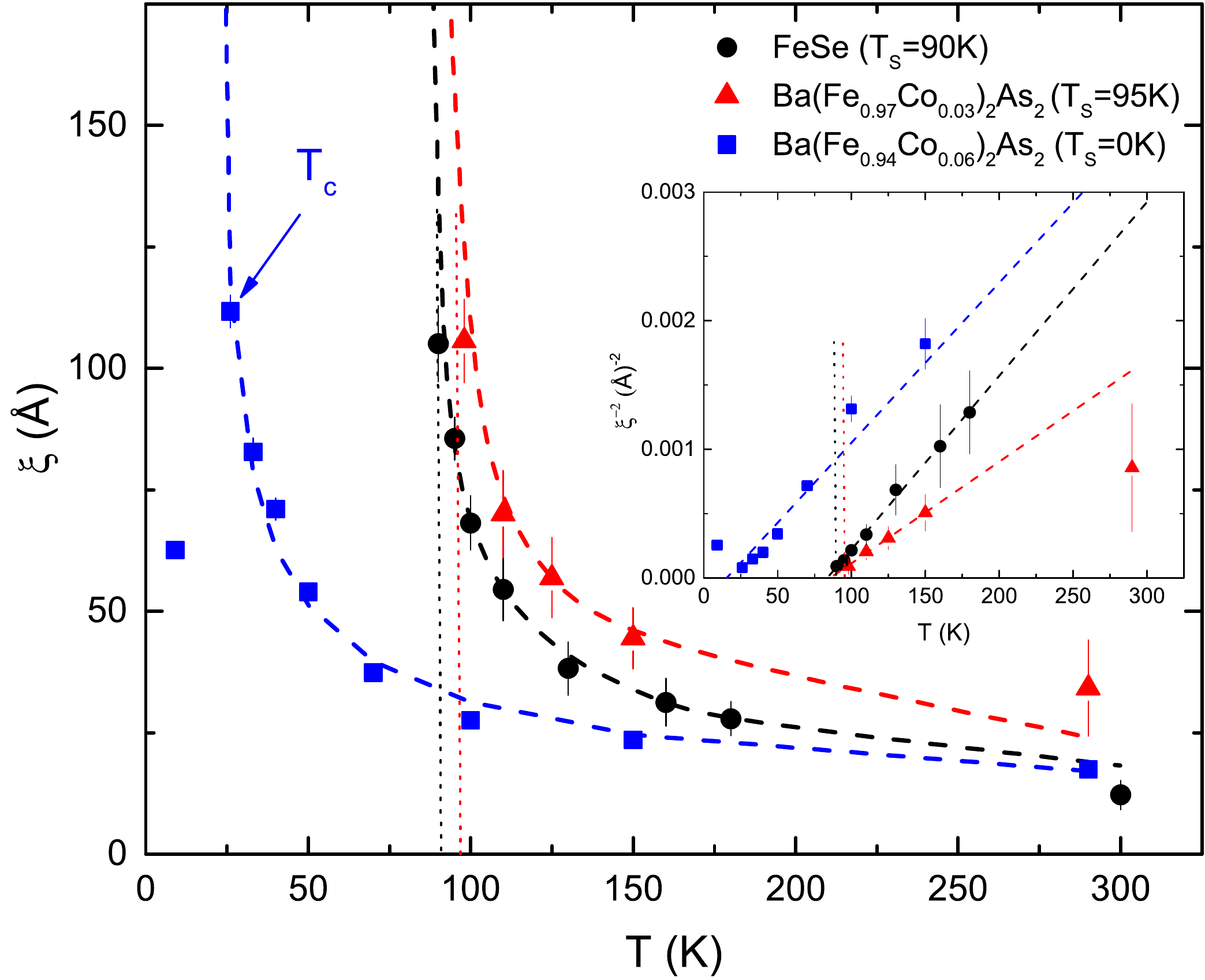}
\caption{Nematic correlation length $\xi$ as a function of temperature for FeSe (black circles), for UD Ba-122 (red triangles), and for OP Ba-122 (blue squares). The dashed lines are power-law fits of the form $\xi=\xi_0/(T-T_0)^{1/2}$. Inset: $\xi^{-2}$ for the materials as in the main panel, with linear fits (dashed lines).}
\label{fig:Xsi_Plot}
\end{figure}

To circumvent this issue, we use the shear modulus reported for highly overdoped Ba(Fe$_{0.745}$Co$_{0.245}$)$_2$As$_2$ in Ref. \cite{Yoshizawa_StructuralQuantumCriticality-2012} to fix the bare shear modulus, $C^0_{66}$, for the UD Ba-122 sample. Thus, we are assuming that the highly overdoped sample does not manifest significant nematic fluctuations at any temperature. We take the renormalized shear modulus, $C_{66}$, from the 3.7\% Co-doped sample reported in the same reference, which is the closest doping level to our sample reported in Ref. \cite{Yoshizawa_StructuralQuantumCriticality-2012}. 

Because for FeSe there is, to our knowledge, no equivalent sample from which to estimate the bare shear modulus, we use the same bare shear modulus as that of Ba(Fe$_{0.745}$Co$_{0.245}$)$_2$As$_2$ in Ref. \cite{Yoshizawa_StructuralQuantumCriticality-2012} and fix the ratio $C_{66}(250\text{K})/C^0_{66}(250\text{K})$ by taking the reported $C_{66}(250\text{K})$ data for Ba(Fe$_{0.963}$Co$_{0.037}$)$_2$As$_2$. To extend this to the general $C_{66}(T)$ for FeSe, we use the reported $Y_{[110]}(T)/Y_{[110]}(250\text{K})$ data on FeSe presented in Ref. \cite{Bohmer_OriginTetragonaltoOrthorhombicPhase-2015} (see Ref. \cite{Weber_Softphononsreveal-2018} for a discussion of the relation between $Y_{[110]}$ and  $C_{66}$). This procedure fixes both the bare and the renormalized shear modulus at all temperatures above $T_S$. Thus, all parameters are fixed except the nematic correlation length $\xi$.

Fig. \ref{fig:Dispersion_Fit} shows the fitted phonon dispersion (solid lines) in (a) UD Ba-122 at $290$K and $98$K ($T_S=95$K) and (b) in FeSe at $300$K and $95$K ($T_S=90$K). The dashed line represents $E(q) = f(q) \sqrt{C_{66}/\rho}$ obtained by setting $\xi=0$. It extrapolates the dispersion at low $q$ and demonstrates the correspondence between the shear modulus and the low-$q$ phonon dispersion (black solid lines in Fig. 2). The dotted line corresponds to zero coupling between the atomic lattice and the electronic degrees of freedom (i.e. by setting $\lambda=0$, $E(q) = f(q) \sqrt{C^{0}_{66}/\rho}$).

The fitted values for nematic correlation length $\xi$ above the superconducting transition temperature, $T_c$, shown in Fig. \ref{fig:Xsi_Plot} are nonzero already at high temperature and rapidly increase on approach to the structural transition. A power-law fit for $\xi$ vs. $T$ using $\xi=\xi_0/(T-T_0)^{1/2}$ yields values of $T_0=84\pm1$ K for FeSe, $T_0=86\pm2$ K for UD Ba-122 and $T_0=20\pm1$ K for OP Ba-122 (Fig. \ref{fig:Xsi_Plot}). Note that only the data above $T_c$ were fit for OP Ba-122, since the increase in nematic correlation length on cooling is reversed by superconductivity \cite{Weber_Softphononsreveal-2018}. The inset in Fig. \ref{fig:Xsi_Plot} demonstrates the universal power-law behavior with the $x$-intercepts at $84\pm1$ K for FeSe, $86\pm1$ K for UD Ba-122 and $16\pm4$ K for OP Ba-122. A fit allowing the exponent $\nu$ to vary freely yields values of $\nu=0.58\pm0.06$ for  FeSe, $\nu=0.40\pm0.04$ for UD Ba-122 and $\nu=0.57\pm0.07$ for OP Ba-122, with no significant effect on the fit quality or values of $T_0$, thus we fixed $\nu$=1/2. The values of $\nu$ of the free fit are all close to 1/2 considering the experimental uncertainty. Note that the total uncertainty for FeSe should be greater than the purely statistical error in the fit, because of additional uncertainty in the unrenormalized dispersion as discussed above.

Our results have important implications. Previous measurements of the uniform nematic susceptibility $\chi_{\mathrm{nem}}$ via elasto-resistance \cite{Fisher16,Coldea15}, Raman spectroscopy \cite{Gallais13,Blumberg16}, NMR \cite{Kissov17}, and elastic moduli \cite{Yoshizawa_StructuralQuantumCriticality-2012,bohmer_nematic_2014} in a variety of different compounds reported a Curie-Weiss behavior $\chi_{\mathrm{nem}} \sim \left(T-T_{\mathrm{CW}}\right)^{-\gamma}$, with a Curie-Weiss temperature $T_{\mathrm{CW}}$ close to the actual structural transition temperature $T_S$ and $\gamma = 1$. Although that behavior is consistent with a mean-field transition, the character of the transition can only be established by probing a second independent critical exponent. Our measurements in two different families of iron-based compounds and at different regimes (underdoped and optimally doped) reveal a clear power-law behavior $\xi \sim (T-T_0)^{-\nu}$, with $T_0$ very close to $T_S$ and $\nu=1/2$. Although the precise determination of actual critical exponents would require careful measurements over a few temperature decades near $T_S$, this set of results suggest that over a wide temperature range the two independent critical exponents $\gamma$ and $\nu$ are those of a mean-field critical point.


We obtain $T_0>0$ at optimal doping, which means that the quantum critical point where $T_0$=0 would be actually at a somewhat higher doping. This is consistent with the previously observed back-bending of the $T_S$ transition line inside the superconducting dome \cite{Nandi10}. This behavior is analogous to copper oxide superconductors where the quantum critical point appears in the overdoped part of the phase diagram (see \cite{Tallon2019} and references therein). 

Since the nematic order parameter is Ising-like \cite{Fernandes10}, it is interesting to understand why the mean-field behavior extends over such a wide temperature range above $T_S$, without seemingly crossing over to an Ising critical behavior. Recent theoretical investigations suggest that the reason is the coupling to the lattice -- more specifically, to the acoustic phonons \cite{Schmalian16,Paul17,Carvalho19}. In real space, these modes mediate long-range interactions between the Ising-nematic degrees of freedom, similarly to the dipolar interaction between Ising spins in a ferromagnet. Such long-range interaction effectively lowers the upper critical dimension of the problem \cite{Cowley76}, rendering the Ising transition mean-field like even in three dimensions. 

Therefore, our observations highlight the key role played by the nemato-elastic coupling, which not only changes the character of the nematic transition, but also extends the impact of the nematic fluctuations to rather high temperatures above $T_S$. Such a coupling has been proposed to be detrimental to the enhancement of $T_c$ by quantum critical nematic fluctuations \cite{Labat17}. Whether this explains the observed behavior of $T_c$ across the phase diagram of chemically-substituted FeSe$_{1-x}$S$_x$, which shows no sizable enhancement upon crossing the putative nematic quantum critical point \cite{Coldea17}, is an interesting topic for future investigation. Moreover, the similar behavior of the nematic correlation length that we observe in FeSe and Ba-122 raises important questions about the interplay between nematicity and magnetism. Although FeSe displays no long-range magnetic order, a strong fluctuating magnetic moment, comparable to that of Ba-122, is observed experimentally \cite{Wang2016}. Whether this is enough to explain the similar behavior of $\xi$ in both compounds is an issue that deserves further studies.

\acknowledgments
D.R. would like to thank I.I. Mazin for helpful discussions. A.M.M. and D.R. were supported by the U.S. Department of Energy (DOE), Office of Basic Energy Sciences, Office of Science, under Con- tract No. DE-SC0006939. Theory work (RMF) was supported by the DOE Office of Science, Basic Energy Sciences, under Award No. DE-SC0020045.  This research used resources of the Advanced Photon Source (APS), a DOE Office of Science User Facility operated for the DOE Office of Science by Argonne National Laboratory under Contract No. DE-AC02-06CH11357. Experimental work was carried out at BL43LXU of the RIKEN SPring-8 Center and at Sector 30 (HERIX) of the APS. 

\bibliographystyle{apsrev4-1}
\bibliography{bibliography}

\end{document}